\documentclass{jps-cp}
\usepackage{txfonts}

\usepackage{bbm}
\usepackage{color}

\title{
  Pion Structure in a Nuclear Medium}

\author{Parada T. P. \textsc{Hutauruk}$^{1}$, Yongseok \textsc{Oh}$^{2,1}$ and Kazuo \textsc{Tsushima}$^{3,1}$}

\inst{$^{1}$Asia Pacific Center for Theoretical Physics, Pohang, Gyeongbuk 37673, Korea \\
  $^{2}$Department of Physics, Kyungpook National University, Daegu 41566, Korea\\
  $^{3}$Laborat\'orio de F\'isica Te\'orica e Computacional-LFTC, Universidade Cruzeiro 
  do Sul, 01506-000, S\~ao Paulo, SP, Brazil
}

\email{parada.hutauruk@apctp.org}
\recdate{January 18, 2019}

\abst{The structure and electroweak properties of the pion in symmetric
  nuclear matter are presented in the framework of the Nambu--Jona-Lasinio model.
  The pion is described as a bound state of a dressed
  quark-antiquark pair governed by the Bethe-Salpeter equation. For the in-medium current-light-quark properties
  we use the quark-meson coupling model, which describes succesfully the properties of hadron in a nuclear medium.
  We found that the light-quark condensates, pion decay constant and pion-quark coupling constant decrease
  with increasing nuclear matter density. We then predict the modifications of
  the charge radius of the charged pion in nuclear matter.}

\kword{pion form factor, quark-meson coupling model, Nambu--Jona-Lasinio model,
  electroweak properties, symmetric nuclear matter}

\begin{document}
\maketitle
%
%------------------------
\section{Introduction}
%-------------------------------------------------------------------
Pions play special roles in understanding the strong interactions of
quantum chromodynamics (QCD) at low energies~\cite{Ioffe05}. The pion electromagnetic form factor
would provide us with information on the non-perturbative aspects of the internal structure as well as
the dynamics of the quarks and gluons.

The phenomenon of medium modifications is one of the most interesting subjects in nuclear and 
hadron physics such as the EMC effect~\cite{EMC83b}. 
In particular, how the properties of hadrons and their internal structures change by the surrounding 
nuclear environment is one of the important issues, which collects special 
attentions~\cite{BR91} in connection with the partial restoration of 
chiral symmetry in a strongly interacting environment~\cite{BR96}.  
The order parameters of this phenomena are the light-quark condensates in nuclear medium
and their changes in nuclear medium are main driving forces for the changes of the hadron properties in nuclear 
medium. However, finding the clear experimental evidence of partial restoration of
chiral symmetry is still challenging.

In the present work, following the formalism of Ref.~\cite{HCT16,Hutauruk18} we report the electroweak
properties of the pion in symmetric nuclear matter as well as the pion form factor. The in-medium
current-light-quark properties obtained in the quark-meson coupling (QMC) model are used as inputs to study
the properties of the in-medium pions in the NJL model. This report is based upon the recent
article~\cite{HOT19}.

%---------------------------------------------------------------
\section{In-medium current-ligth-quark properties based on the QMC model}
%--------------------------------------------------------------------------
The QMC model~\cite{Guichon88} has been successfully applied to many phenomena of nuclear and hadron 
systems. Recently, studies have been extended to the calculation of
the medium modifications of the nucleon weak and 
electromagnetic form factors on the neutrino mean free path in dense matter~\cite{HOT18}.

In symmetric nuclear matter, the isospin-dependent $\rho$-meson field vanishes in the Hartree 
approximation. The nucleon Fermi momentum $k_F^{}$, nucleon (baryon) density $\rho_B^{}$, and
the scalar density $\rho_{s}^{}$ of nuclear matter are defined as
\begin{align}
  \label{eqintro4}
  \rho_{B}^{} &= \frac{\gamma}{(2\pi)^3} \int d {\bf k} \theta ( k_F^{}  - | {\bf k} | ) = 
  \frac{\gamma k_F^3}{3 \pi^2}, \qquad
  \rho_{s}^{} = \frac{\gamma}{(2\pi)^3} \int d {\bf k}  \theta ( k_F^{} - | {\bf k} | )
  \frac{M_N^{*} (\sigma)}{\sqrt{M_N^{*2} (\sigma ) + {\bf k}^2}},
\end{align}
where $\gamma = 4$ for symmetric nuclear matter and $\gamma = 2$ for asymmetric nuclear matter. 
The Fermi momenta of the proton and neutron $k_F^{p,n}$ are respectively determined by $\rho_p^{}$ and 
$\rho_n^{}$ with $\rho_B^{} = \rho_p^{} +\rho_n^{}$.

In the QMC model, nuclear matter is treated as a collection of the nucleons that are assumed to be 
non-overlapping MIT bags. 
The Dirac equations for the light $q$ ($u$ and $d$) quarks inside the bag are given by
\begin{align}
\label{eqintro5}
\left[ i \gamma \cdot \partial - \left( m_q^{} - V_{\sigma}^{q} \right)  \mp \gamma^{0} V_{\omega}^{q} 
\right] \left( \begin{array}{c} \psi_u(x)  \\ \psi_{\bar{u}}(x) \\ \end{array} 
\right) &= 0 , \qquad
\left[ i \gamma \cdot \partial - \left( m_q^{} - V_{\sigma}^{q} \right) \mp \gamma^{0}  V_{\omega}^{q} 
 \right] \left( \begin{array}{c} \psi_d(x)  \\ \psi_{\bar{d}}(x) \\ \end{array} 
\right) &= 0,
\end{align}
which define the effective current quark mass $m_q^{*} \equiv m_q^{} - V_{\sigma}^{q}$.
Here we assume $m_u = m_d = m_q$.
The scalar and vector mean fields in symmetric nuclear 
matter are defined by
$V_{\sigma}^{q}  \equiv g_{\sigma}^{q} \langle \sigma \rangle$ and $V_{\omega}^{q}
\equiv g_{\omega}^{q}  \delta^{\mu 0} \langle  \omega^{\mu} \rangle$, respectively.

By solving the self-consistent equation for the scalar $\sigma$ mean field, 
the total energy per nucleon is obtained as
\begin{align}
  \label{eq:pionmed12}
  E^{\rm tot}/A &= \frac{4}{(2\pi)^3 \rho_B^{}} \int d {\bf k} \, \theta (k_F^{} - | {\bf k} |) \sqrt{M_N^{*2} (\sigma) 
+ {\bf k}^2}  + \frac{m_\sigma^2\, \sigma^2}{2\rho_B^{}} 
+ \frac{g_\omega^2\, \rho_B^{}}{2 m_\omega^2}.
\end{align}
The coupling constants $g^q_\sigma$ and $g^q_\omega$ are determined by fitting the 
binding energy, 15.7~MeV, of symmetric nuclear matter at the saturation density 
and they are respectively related with $g_\sigma$ and $g_\omega$ by $g_\sigma^{} = 3 S_N(\sigma=0) g^q_\sigma$ and 
$g_\omega^{} = 3 g^q_\omega$, where $S_N(\sigma)$ is defined through~\cite{Guichon88}
\begin{eqnarray}
\dfrac{\partial M_{N}^*(\sigma)}{\partial \sigma}
&=& - 3 g_{\sigma}^q \int_{\rm bag} d\,{\bf y} \ {\overline \psi}_q({\bf y})\, \psi_q({\bf y})
\equiv - 3 g_{\sigma}^q S_{N}(\sigma) = - \dfrac{\partial}{\partial \sigma}
\left[ g^{N}_\sigma(\sigma) \sigma \right],
\label{Ssigma}
\end{eqnarray}
with $\psi_q$ being the lowest mode bag wave function in medium.
These relations determine the in-medium quark dynamics, which is needed to compute the medium modifications 
of \textit{hadron} properties in nuclear medium.
That is, we assume that the in-medium light-quark properties obtained for the nucleon are not much different from those
of light quarks in other hadrons. 
In the present work, therefore, we will explore the in-medium pion properties
with the modified quark properties determined by the 
in-medium nucleon properties. 
%

%--------------------------------------
\section{In-medium pion properties}
\begin{figure}[t]
  \centering
  \includegraphics[width=0.45\columnwidth]{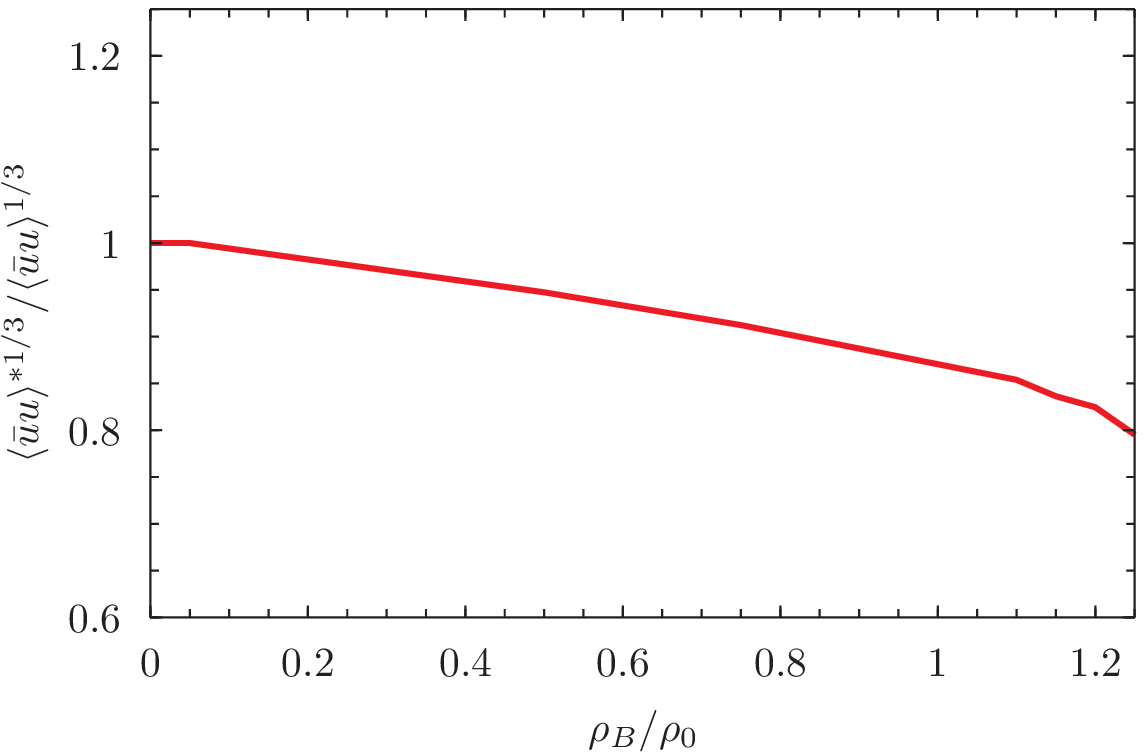} \hspace{0.1cm}
  \includegraphics[width=0.45\columnwidth]{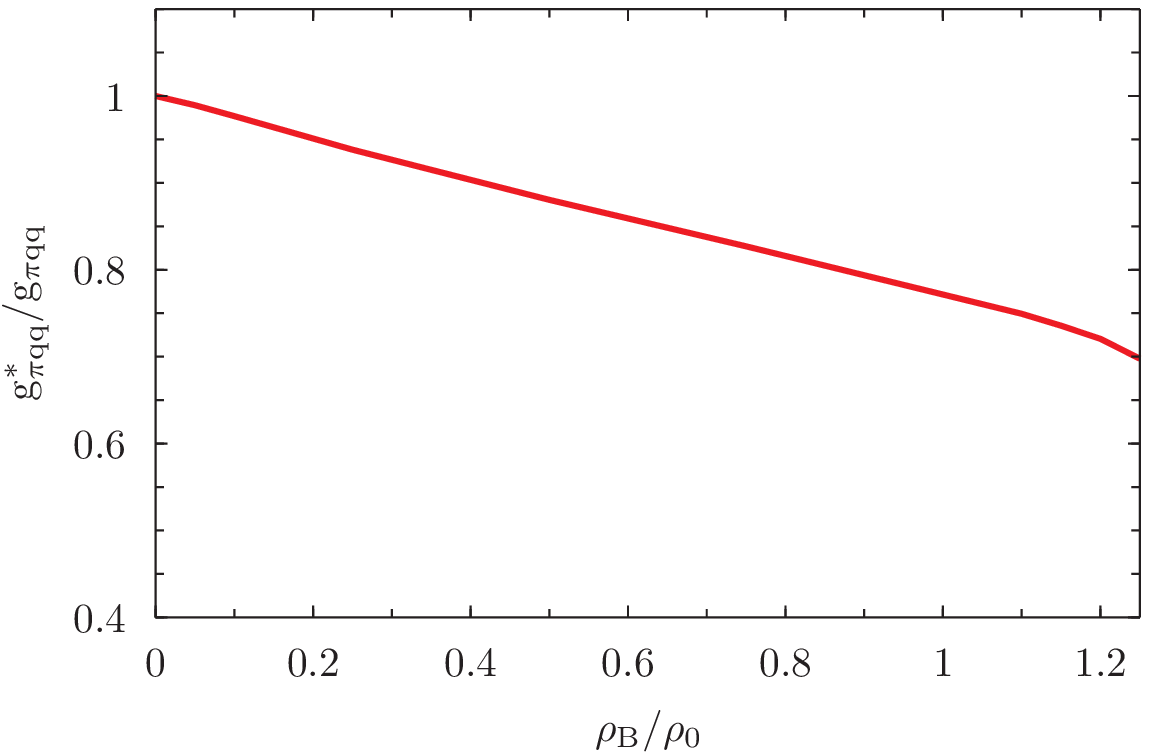}
  \caption{\label{fig4} 
    The ratio of the in-medium to vacuum quark condensate as a function of $\rho_B^{} / \rho_0^{}$ (left panel).
    The ratio of the in-medium to vacuum pion-quark coupling constant
    as a function of $\rho_B^{} / \rho_0^{}$ (right panel).}
\end{figure}

\begin{figure}[t]
  \centering
  \includegraphics[width=0.45\columnwidth]{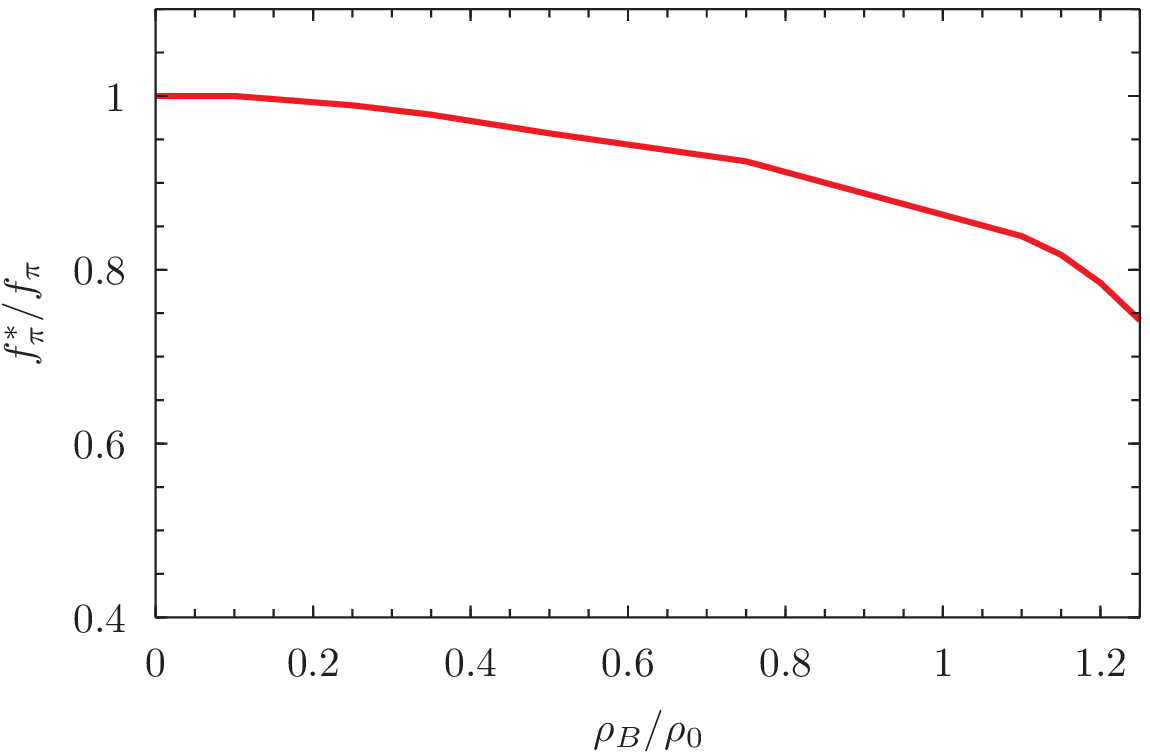}\hspace{0.1cm}
  \includegraphics[width=0.45\columnwidth]{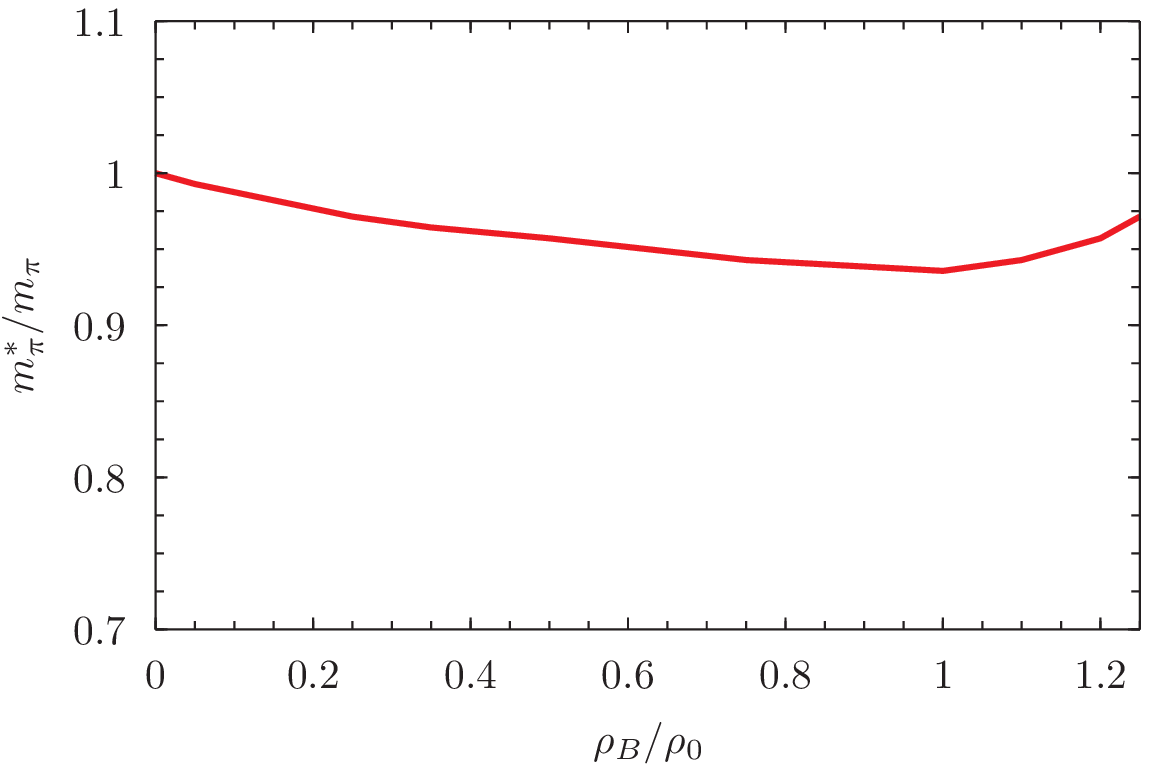}
  \caption{\label{fig6} 
    The ratio of the in-medium to vacuum pion decay constant as a function of $\rho_B^{} / \rho_0^{}$ (left panel).
    The ratio of the in-medium to vacuum pion mass as a function of $\rho_B^{} / \rho_0^{}$ (right panel).} 
\end{figure}

%---------------------------------------------------------------------------
Being equipped with the NJL-model formalism and QMC model for the medium-modified current-light-quark properties, 
we compute the pion properties in nuclear medium.
Using the in-medium current-light-quark properties obtained in the QMC model with $m_q = 16.4$~MeV,
we calculate the effective constituent quark mass $M^*_u$, in-medium pion mass, in-medium pion decay constant,
in-medium quark condensate, and in-medium $\pi q q$ coupling constant.
The results are illustrated in Figs.~\ref{fig4}--\ref{fig6} as functions of $\rho_B^{} / \rho_0^{}$
with $\rho_0$ being the normal nuclear density, 0.15 fm$^{-3}$.
%
%%%  FIG 4
%%%%%%%%%%%%%%%%%%

%%%  FIG 6
%%%%%%%%%%%%%%%%%%
The left panel of Fig.~\ref{fig4} shows the ratio of the in-medium to vacuum quark condensates.
We found that this ratio decreases as the nuclear matter density increases. In the right panel of Fig.~\ref{fig4}
we found that the ratio of the in-medium to vacuum pion-quark coupling constant decreases with increasing
nuclear matter density. Our results for the ratio of the in-medium to vacuum pion decay constant are
presented in the left panel of Fig.~\ref{fig6}. Again, the ratio is found to decrease as
the density increases. The medium modifications on the pion mass is shown in the right
panel of Fig.~\ref{fig6}. Here we confirm that the pion mass is almost constant up to
1.25~$\rho_0$. This justifies our assumption that $m_\pi^* \approx m_\pi$ up to about
normal nuclear density.

%------------------------------------------------
\section{In-medium electromagnetic form factors}
%-------------------------------------------------------------------------------
We calculate the electromagnetic form factor of the positively charged in-medium pion.
The in-medium pion electromagnetic form factors are determined by adopting 
the method of Refs.~\cite{HCT16,Hutauruk18} with a dressed quark-photon vertex.
The in-medium electromagnetic form factor is obtained by modifying the quark propagator as  
\begin{equation}
  \label{eq:prognjl}
  S_q^{*}(k^{*}) =  \frac{1}{ k\!\!\!/^{*} - M_q^{*} + i \epsilon },
\end{equation} 
where $M_q^{*}$ is the effective constituent quark mass and the in-medium quark momentum is given   
by $k^{*}_\mu = k_\mu + V_\mu$ with vector field $V_\mu=(V_0,{\bf 0})$ by neglecting the modifications 
of the space component of the quark momentum that is known to be small~\cite{KTT98}.
In the form factor calculation, the vector field, which enters the propagator in Eq.~\eqref{eq:prognjl},
can be eliminated by the shift of the variable in the integration~\cite{DTERF14}.
The in-medium form factor of the pion is then given by
\begin{align}
\label{eq:bareKplus}
F_{\pi^{+}}^{\text{* (bare)}}(Q^2) &= \left( e_u - e_d \right) f^{* \ell \ell}_\pi (Q^2) ,
\end{align}
where $\ell = u, d$, and $e_{u/d}$ are the $u/d$ quark charges.
The superscript ``(bare)'' in Eq.~(\ref{eq:bareKplus})
means that the quark-photon vertex is elementary, that is,  
$\Lambda_{\gamma q}^{\mu\text{(bare)}} = \hat{Q}\,\gamma^\mu$, where $\hat{Q}$ is the quark charge operator. 
The first superscript $a$ in $f_\pi^{* ab}(Q^2)$ indicates the struck quark 
and the second superscript $b$ means the spectator quark (See Ref.~\cite{HOT19} for the details).
%
%%%  FIG 10
\begin{figure}[t]
  \centering
  \includegraphics[width=0.4\columnwidth]{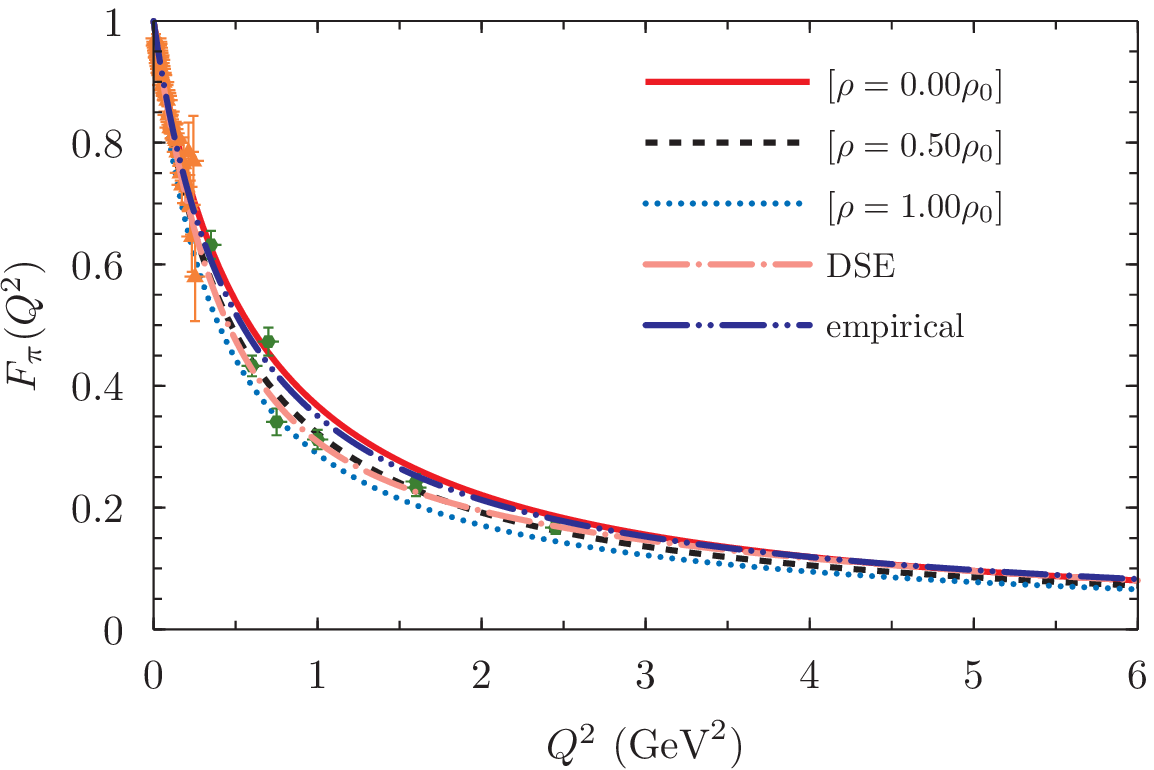} \hspace{0.1cm}
  \includegraphics[width=0.4\columnwidth]{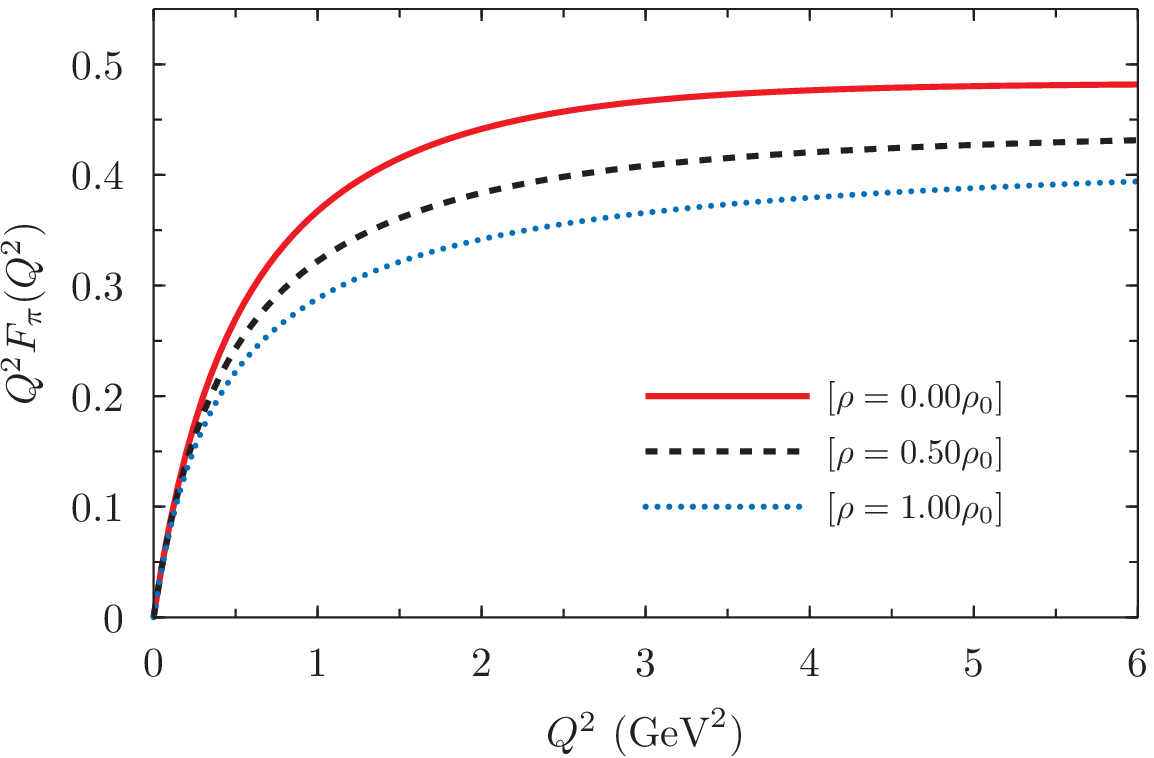}
  \caption{\label{fig10} 
    The electromagnetic form factor of positively charged pion for different nuclear matter densities (left panel).
    The solid, dashed, and dotted lines are for $\rho_B^{} / \rho_0^{} = 0.0$, $0.5$, and $1.0$, respectively. 
    The solid line is to be compared with the DSE predictions of Ref.~\cite{CCRST13},
    the empirical parameterization of Ref.~\cite{ABBB84},
    and the experimental data from Refs.~\cite{NA7-86,ABBB84,JLAB_Fpi-08}.
    Results for $Q^2 F_{\pi}^{*} (Q^2)$ for various nuclear matter densities (right panel).}
\end{figure}
%%%%%%%%%%%%%%%%%%
%
Our numerical result for the elastic form factors of the positively charged pion in medium
are shown in the left and right panels of Fig.~\ref{fig10}. The left panel of Fig.~\ref{fig10}
shows that the in-medium pion electromagnetic form factor is suppressed with increasing density.
Shown in the right panel of Fig.~\ref{fig10} are the same results as in the left panel of Fig.~\ref{fig10}
but for $Q^2 F_\pi (Q^2)$. The medium effects on the suppression of the pion form factor are clearly
seen and it reduces by about 10\% at the normal nuclear density. Our calculations show that the charge
radius of the in-medium pion increases with density (See Ref.~\cite{HOT19} for details on the in-medium
charge radius prediction).

%-------------------------------------
\section{Summary}
%---------------------------------------------------------------
In summary, we have reported the electroweak properties and the structure of the pion in symmetric
nuclear matter in the framework of the NJL model. The in-medium modifications of the current-light-quark properties
are calculated by the QMC model and are used as inputs of the NJL model calculations.
We found that the electroweak properties of the pion, namely, the ratios of in-medium to vacuum
quark condensates, pion-quark coupling constant and the pion decay constant decrease with increasing
the nuclear matter density. We also found that the elastic form factor of the in-medium pion is suppressed compared with
the form factor in vacuum as the nuclear density increases. Consequently, the charge radius
of the charged pion increases as the nuclear matter density increases.

\end{document}